\documentclass[aps,pre,twocolumn]{revtex4-1}
\usepackage{graphicx,amssymb,amsmath,epsf,color,hyperref}

\begin{document}
\title{Spatial uniformity in power-grid system}
\author{Mi Jin Lee}
\thanks{Present address: Department of Physics, Inha University, Incheon 22212, Republic of  Korea }
\author{Beom Jun Kim}
\email{Corresponding author: beomjun@skku.edu}
\affiliation{Department of Physics, Sungkyunkwan University, Suwon 16419, Republic of Korea}
\begin{abstract}
Robust synchronization is indispensable for stable operation of a power grid.
Recently, it has been reported that a large number of decentralized generators,
rather than a small number of large power plants, provide enhanced
synchronization together with greater robustness against structural
failures.  In this paper, we systematically control the spatial uniformity of the
geographical distribution of generators, and conclude that the more uniformly
generators are distributed, the more enhanced synchronization occurs.  In the
presence of temporal failures of power sources, we observe that spatial
uniformity helps the power grid to recover stationarity in a shorter time.  We
also discuss practical implications of our results in designing the structure
of a power-grid network.

\end{abstract}

\maketitle

\section{Introduction}
\label{sec:intro}
The emergence of synchronization has been an important issue in the statistical physics
community. It has been observed in a variety of phenomena like the circadian
rhythm~\cite{circadian}, epilepsy in the brain~\cite{epil}, and the
London Millennium Bridge~\cite{bridge}. Examples of synchronization phenomena
exist also in social behaviors~\cite{social1, social2} and in the power
grid~\cite{epjb}.  It is to be noted that the emergence of synchronization can be
desirable or not, depending on what is required for a given system to function
properly.  For example, an epileptic seizure in a human brain and a large
vibration of the London Millennium Bridge are caused by undesired
synchronization, and recent research has reported how to inhibit such
unwanted synchronization behavior~\cite{suppress}.  In contrast, for the power
grid, which is the main focus of the present paper, robust synchronization is
an essential ingredient for the system to work in a stable manner.  If a power
plant in the grid is not running at a proper reference frequency, or the phase
angle of the voltage produced by the plant does not match that of the grid, the
plant cannot properly convey electric power to the grid, and it is possible to
result in a short circuit and damage to the generator~\cite{gridsynch1,
gridsynch2}.  A malfunctioning generator can yield cascading failures in the
grid in the worst case, leading to a large-scale blackout.

Synchronization in the power-grid system has been studied from various
points of view. The swing equation, which is similar to the Kuramoto model in
statistical physics, has been suggested to describe the dynamics of the power-grid
system~\cite{epjb, motternjp}.  Stability of the synchronized state is also an
important issue and has been studied through the use of several methods such as
the basin stability~\cite{kruths, hkim1} and the Lyapunov exponent~\cite{msf}.
A power grid is of course a typical example of complex networks. From this
perspective, the impact of the network structure on the synchronous state~\cite{impact}
and the optimal network topology for enhanced synchronization~\cite{optimize}
have been investigated.

An interesting finding in existing studies is that decentralized
generators provide enhanced synchronization together with robustness against
structural failures~\cite{decentral}.  This is a particularly important result
in view of the growing recent interest in renewable energy sources.  If a
large number of small generators based on renewable sources are
connected in the power grid in the form of a complex network, understating
of the effect of decentralized power generators can be very important.

In this work, we focus on how to distribute decentralized power
generators in two-dimensional space to enhance the synchronization of the power grid.
We control spatial uniformity of generators in two ways: First, we divide
the whole two-dimensional area into available
and unavailable regions for the positions of
power plants and study how the change of the available area affects
synchronization. Second, we impose various sizes of superlattice unit cells
on top of the original two-dimensional square lattice. In each superlattice
unit cell, power plants are put near the center of the unit cell. In the limiting
case where one superlattice unit cell covers the whole system, all plants
are put only in the central part of the whole region. As the size of the
superlattice becomes smaller, power plants tend to be spread more uniformly across
the whole system.

The present  paper is organized as follows: In Sec.~\ref{sec:model}, we
introduce the Kuramoto model with inertia to describe the dynamics of the power
grid~\cite{epjb}. We also check how the synchronization behavior depends on
the system size and the number of generators.
In Sec.~\ref{sec:uniformity}, we describe how we control the
spatial uniformity of generators in two dimensions in two different ways, and
explain our results. Section~\ref{sec:summary} is devoted to a discussion and summary.

\section{Equation of Motion for The Power Grid}
\label{sec:model}

We sketch the derivation of the equation of motion for a power grid in
Ref.~\cite{epjb}.  The power grid is composed of two types of nodes, generators and
consumers.  A consumer node in this work may represent a large group of
consumers in reality, like a city.  Energy from a natural source is first
converted to mechanical energy, and an electric generator then converts it to electric energy.  
In a hydroelectric power plant, for instance, the
gravitational potential energy of water in a dam is converted to the rotational
kinetic energy of a turbine,  which generates an alternating current (ac)
electric power via
Faraday's law of electromagnetic induction. In a thermal power plant based
on a fossil energy source, the chemical energy in the fuel is used to boil
the water, and the pressure from the steam provides the mechanical rotational
energy injected into an electric generator.

In this work we denote the
mechanical energy injected into a turbine of the generator at
node $i$ as $P_i$.
The dynamic variable for a generator node in the power grid is the
turbine's rotational phase angle, which is directly related to
the phase of the ac voltage generated from the generator. We assume that
the former equals the latter and write it as $\theta_i$ for the
generator node $i$.
In the research community of the power grid,
the consumer node $j$ is also assigned the phase variable $\theta_j$ in
the same way as for the generator node. One can assume that each consumer
has an electric motor to convert the supplied electric power into mechanical
power, and $\theta_j$ is either the rotational phase angle of the motor
or the phase of the voltage for the consumed power.
For both the generator and the consumer nodes, the instantaneous
phase at the $i$th node is written as $\theta_{i}(t)=\Omega t+\phi_{i}(t)$,
where $\Omega = 2\pi f$ with a grid reference frequency $f = 60$ or 50 Hz,
and $\phi_i(t)$ measures the deviation from the rotating reference frame.

The injected mechanical energy is then transformed to the rotational kinetic
energy of the turbine (the motor) of the generator (the consumer) node and the
electric energy to be transmitted to other nodes in the power grid. Of course,
some part of the injected energy is lost in the form of dissipation due to
friction.  Accordingly, one can write the energy (or the power) conservation condition  
in the form
\begin{equation}
\label{eq:balance}
P_i  = P_{\textrm{kin}, i}+P_{\textrm{diss}, i}+\sum_{j}P_{\textrm{trans}, ij},
\end{equation}
where $P_{\textrm{kin}}$ and $P_{\textrm{diss}}$ are powers corresponding to
the rotational kinetic energy and the dissipated energy, respectively,
and the last term is the power transmitted from $i$ to other connected nodes in
the power grid.
The power, the energy per unit time, for the rotational kinetic energy is given by
$P_{\textrm{kin}, i}=\frac{d}{dt}\left( \frac{1}{2}I\dot{\theta}_i^2 \right)$ with
the moment of inertia $I$ of the turbine or the motor.
The power for the dissipated energy is written as
$P_{\textrm{diss}, i}=\kappa \dot{\theta}_i^2$ with the dissipation coefficient $\kappa$.
The power transmitted from $i$ to $j$ through a transmission line is
written as
$P_{\textrm{trans}, ij} =P_{\textrm{max}}A_{ij}\sin(\theta_i-\theta_j)$, where $P_{\textrm{max}}$ is
the maximum allowed value of the transmitted power, and $A_{ij}$ is the element of
the symmetric adjacency matrix, i.e., $A_{ij}=A_{ji}= 1$ if $i$ and $j$ are
connected, and $A_{ij}=A_{ji} = 0$ otherwise. (See Appendix~\ref{App:flow} for
the calculation of the transmitted power.) We remark that although the present work
focuses on a two-dimensional square grid with only geographically local couplings, it is straightforward
to apply our framework to a general adjacency matrix which may describe
very long transmission cables in the power grid.

Under the realistic assumption that the power grid works almost at the reference angular
frequency $\Omega$, i.e., $\dot\theta_i =  \Omega + \dot\phi_i \approx \Omega$ and thus
$|\dot{\phi_i}|\ll\Omega$, we obtain
$P_{\textrm{kin}, i}\approx I\ddot{\phi_i}\Omega$ and
$P_{\textrm{diss}, i}\approx  \kappa\Omega^{2}+2\kappa\Omega\dot{\phi_i}$.
Using $\theta_j-\theta_i=\phi_j-\phi_i$,
Eq.~(\ref{eq:balance}) is then written as
\begin{equation}
\label{eq:balance2}
\ddot{\phi_i} = \bar{P}_{i} - \alpha
\dot{\phi_i}+K\sum_{j=1}^{N}A_{ij}\sin(\phi_j-\phi_i),
\end{equation}
where $N$ is the size of the power grid,
$\bar{P}_{i} \equiv (P_{i}-\kappa\Omega^2)/I\Omega$, $\alpha \equiv 2\kappa/I$, and
$K \equiv P_{\textrm{max}}/I\Omega$. It is to be noted that a power generating node
can have a negative value of the net power ($\bar P_i < 0)$ if the injected mechanical power
is less than the rotational kinetic power.
Henceforth we regard a node as a generator (a consumer) if $\bar{P}_{i} > 0$ ($\bar{P}_{i} < 0$).

The power conservation equation,~(\ref{eq:balance2}), for the stationary state $\dot\phi_i = \ddot\phi_i = 0$
is written as $0=\bar{P}_{i}+K\sum_{j}A_{ij}\sin(\phi_j-\phi_i)$, and thus we get the power balance condition
$\sum_i\bar{P}_{i} = 0$ from the symmetry $A_{ij} = A_{ji}$ of the adjacency matrix. It is important to note
that if the conservation of the total power, i.e., $\sum_i \bar{P}_i= 0$, is violated,
the system cannot have a stationary state.
In the present work, we assume that all generator nodes have equal power
generating capacity (except in Appendix B where we discuss the case for a few
large power plants at the boundary) and 
all consumer nodes consume an equal amount of power. 
In other words, we use $\bar{P}_{i}=cP$ for a generator node, and
$\bar{P}_{i}=-P$ for a consumer node, respectively, with $P>0$.  The positive
constant $c$ is simply determined from $\sum_{i}\bar{P}_{i}=0$.  Note that as
the number of generators $N_{\rm gen}$ is decreased the power provided by each
power plant should increase, i.e., $c$ is a decreasing function of $N_{\rm
gen}$ since $c = N/N_{\rm gen} - 1$. Since the population density in the real world
is well-known to be nonuniform, the above assumption that all consumer nodes
have the same value of power consumption cannot be valid in reality.
Nevertheless, we emphasize that our framework can easily be modified once
the power consumption of each consumer node is determined from the real data. 
Similarly to Ref.~\cite{braess}, we set $\alpha \equiv 1$ and $P \equiv 1$ for convenience.

We use the initial condition $\phi_i(t=0) = 0$ and numerically integrate the
equation of motion in Eq.~(\ref{eq:balance2}) by using the second-order Runge-Kutta
algorithm~\cite{Press} with the discrete time step $\Delta t = 0.001$ till $t=T$ is approached.
For given constraints such as the system size $N$, the number of generator nodes $N_{\rm gen}$,
and the available region for them,  we randomly choose the generator nodes among
available nodes on a two-dimensional square lattice (remaining nodes become consumer nodes).
Each sample with a random realization of generator nodes is evolved in time, and
it is observed that some samples eventually approach stationary states and others do not.
We check how long we need to wait to judge whether or not the system eventually arrives at the stationary
state and find that $T=150$ and $T=200$ do not make any difference. In other words,
if the system arrives at a stationary state, it does so well before $T=150$. If the system
does not settle down to a stationary state before $T=150$, it remains in a nonstationary state
even after $T=200$.
As a key quantity to measure the global synchrony at a given coupling constant $K$,
we use the conventional order parameter $r(t)$ for the Kuramoto model,
\begin{equation}
r(t) \equiv \frac{1}{N}\left| \sum_j e^{i\phi_j(t)} \right|.
\end{equation}
If the system does not arrive at the stationary state till $t=T(=200)$, we
suppose that the system will be nonstationary to eternity and assign the
value $m=0$ for our order parameter $m$.
If the system approaches the stationary state much earlier than $T=200$, we assign $m = r(T=200)$.
We take the ensemble average over 200 different realizations of generator nodes
to get $\langle m \rangle$. The logic behind the above
definition of the order parameter is that $m$ must have a nonzero value
for the power grid to function properly. If the power grid keeps fluctuating without
settling down to the stationary state, our definition of $m$ indicates failure
of the power grid. Otherwise, if the power grid approaches a stationary state
but the level of synchrony is not good enough, our definition of $m$
also indicates failure of the power grid.


\begin{figure}
\includegraphics[width=0.45\textwidth]{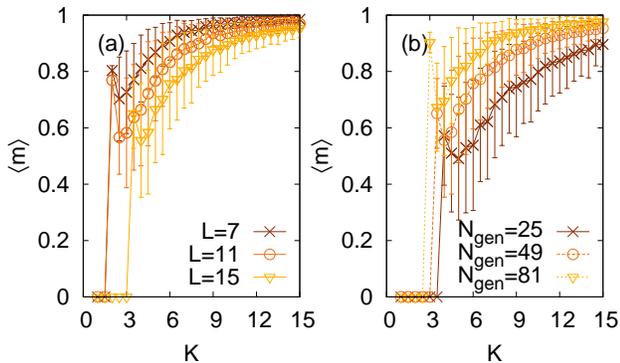}
\caption{(Color online) Ensemble average of the order parameter $\langle m \rangle$
versus coupling strength $K$ when power generators are randomly placed on a two-dimensional
square lattice of the size $N=L\times L$. (a) We fix the generator density $N_{\rm gen}/N \approx 0.2$
and vary the system size. ($N_{\rm gen} = 9, 25,$ and $49$ for $L=7, 11,$ and $15$, respectively.)
(b) We fix the system size $L=15$ and vary the number of generators $N_{\rm gen} = 25, 49,$ and $81$.
From (a) and (b),
we conclude that the level of synchrony is a decreasing function of the system size $L$
and an increasing function of the number of generators $N_{\rm gen}$. }
\label{fig:LNgen}
\end{figure}

In Fig.~\ref{fig:LNgen}(a), we show the ensemble average (over 200 samples)
of the order parameter $\langle m\rangle$
versus the coupling strength $K$ for various sizes $N=(L\times L)$ of two-dimensional square lattices
at almost the same value
of the generator density $N_{\textrm{gen}}/N$: $N_{\rm gen} = 9,  25,$ and $49$
are used for $L=7,11,$ and $15$ (and thus the generator density $N_{\textrm{gen}}/N \approx 0.2$).
Positions of generators are randomly picked in the whole system.  In
Fig.~\ref{fig:LNgen}(a), it is clearly shown that as the system size is
increased, $\langle m \rangle$ is decreased at any value of $K$, which is in
accord with the known result that the locally coupled oscillators in two
dimensions are not synchronized at any finite value of $K$ in the thermodynamic
limit~\cite{localcouple}.  We thus expect that if $L$ is increased further
$\langle m \rangle \rightarrow 0$ for any $K$.  However, a power grid in the
real world is always of a finite size, and we are interested only in how to
distribute generators for a given system size.

In Fig.~\ref{fig:LNgen}(b), we plot $\langle m \rangle$ versus $K$ for
$N_{\textrm{gen}} = 25, 49,$ and $81$ at the fixed system size of $L=15$.
One can understand the observed behavior that the same level
of $\langle m \rangle$ is achieved at a lower value of $K$ when the number of
generators is larger as follows: The coupling constant $K$ in our model plays
the role of the capacity of the power transmission cable. Consequently,
as the number of generators becomes larger, the power that each generator
should supply becomes less, and thus we can achieve a sufficient level of synchrony
with a smaller value of $K$.
From the above investigations in Fig.~\ref{fig:LNgen}(a) and ~\ref{fig:LNgen}(b),
the roles played by the number of generators and the size of the system
can be summarized in a simple manner: The level of synchrony is reduced for
a larger system size and for fewer generators.
Accordingly, henceforth, we fix the system size  and
the number of generators and focus on the effect of
the pattern of the spatial distribution of generators.
Also note that our result in Fig.~\ref{fig:LNgen}(b) is consistent with
the finding that decentralized generators exhibit better synchrony~\cite{decentral}.

\begin{figure}
\includegraphics[width=0.45\textwidth]{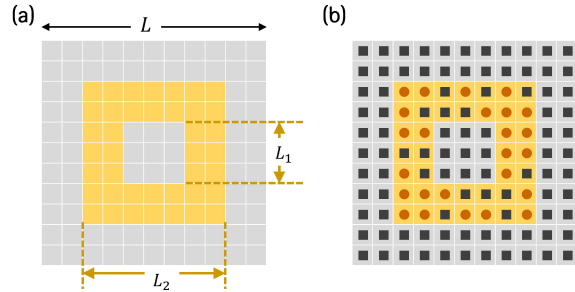}
\caption{(Color online) Method 1: Control of spatial uniformity by changing
the available region for generators. The
two-dimensional square lattice of linear size $L$ is divided into two
regions (gray and yellow). The yellow region between the outer square of 
size $L_2$ and the inner square of size $L_1$ is allowed for generator
locations.  Among all $S = L_2^2 - L_1^2$ points in the available yellow 
region, $N_{\rm gen}$ points are randomly selected as positions of generators.
(b) An example realization of generator positions (brown circles). Small black
squares are the locations of consumers.}
\label{fig:method1}
\end{figure}

\section{Spatial Uniformity and Synchronization}
\label{sec:uniformity}

In this section, we propose two methods to control the spatial uniformity of
generators in a power grid in two dimensions: First, we define the region in
which generators can be located as the outside of the square of the linear size
$L_1$ and the inside of the square of the linear size $L_2$. The centers of both
squares coincide with the central position of the whole two-dimensional power grid.
As $L_1$ and $L_2
(>L_1)$ are changed (and so is the available area $S \equiv L_2^2
-L_1^2$) the degree of spatial uniformity  of the generator distribution is
systematically altered. Second, we place a superlattice structure of linear
size $l$ on top of the underlying square lattice, and put a given number of
generators in the central region in each superlattice unit cell. Generators are spread
more uniformly for small $l$, and in the limiting case where $l$ equals the
linear size $L$ of the whole system the spatial uniformity is the lowest. We
describe our results for the first method (using $L_1$ and $L_2$) of
changing the available region in Sec.~\ref{subsec:spadist} and
for the second method (using various sizes $l$ of
superlattices) in Sec.~\ref{subsec:super}.

\subsection{Method 1: Change of available region}
\label{subsec:spadist}

Figure~\ref{fig:method1}(a) illustrates how we control the spatial distribution
of generators by changing the region available for generator locations. Note
that the power-grid network in the present work has the structure of a
conventional square lattice of size $N = L \times L$ in two dimensions,
with the open boundary condition. We believe that use of the open boundary
condition rather than the  periodic boundary condition makes much more sense
since most power plants in reality often provide electric power 
only within a country. 
In our model system, each node which is not on the boundary in
the power grid has four nearest neighbors with equal degree $k=4$.
We use two squares of sizes $L_1$ and $L_2$, and only the region
between the inner ($L_1$) and the outer ($L_2$) squares
[yellow region in Fig.~\ref{fig:method1}(a)] is
allowed for the locations of generators. The allowed region has width $w \equiv (L_2 - L_1)/2$
and area $S \equiv L_2^2 - L_1^2$. (Note that $0\leq L_1 < L_2 \leq L$.)
In the allowed region of the area $S$, we put $N_{\rm gen }$ generators
at randomly chosen positions ($N_{\rm gen} \leq S$).
As an example, one random realization of generator locations is depicted
in Fig.~\ref{fig:method1}(b) for $L=11$, $N_{\textrm{gen}}=25$, $L_1=3$, and $L_2=7$
(and thus $w = 2$ and $S = 40$).

\begin{figure}
\includegraphics[width=0.45\textwidth]{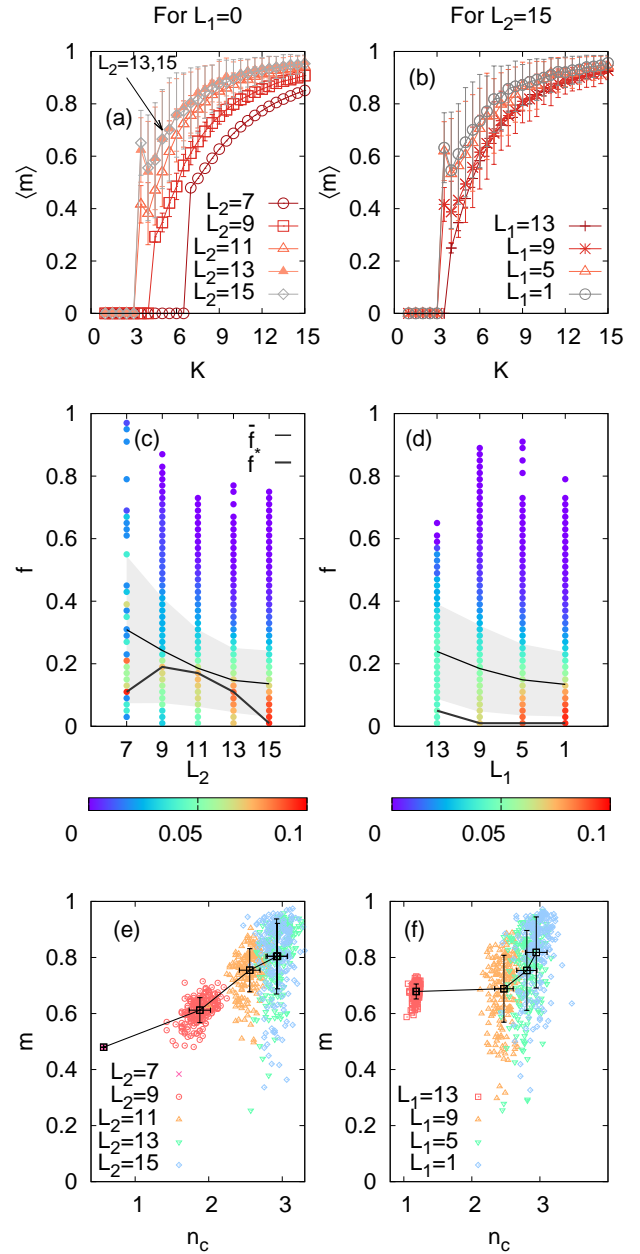}
\caption{(Color online) We change the spatial uniformity
as depicted in Fig.~\ref{fig:method1} for a given number of generators
$N_{\rm gen} = 49$ for a power grid of size $L=15$.
The synchronization order parameter $\langle m \rangle$ versus the coupling
constant $K$ (a) for various values of $L_2$ at fixed $L_1 = 0$ and (b) for
various values of $L_1$ at fixed $L_2=L=15$.
(c, d) Normalized histograms $P(f)$  for the
power transmitted through each link at $K=7$, corresponding to (a) and (b),
respectively, are displayed as a color map. The lines for $\bar f \equiv \sum_f
f P(f)$ and for $f^* \equiv \arg\max P(f)$ are also shown.
It is clearly shown
that as generators are more uniformly spread in space [as $L_2$ is increased in
(c) and as $L_1$ is decreased in (d)], $f$ tends to have small values.
(e, f) Scatter-plots of $m$ versus $n_c$ [see Eq.~(\ref{eq:pair})] at $K=7$, corresponding to (c) and (d),
respectively.
}
\label{fig:L1L2}
\end{figure}

In order to check the effect of spatial uniformity for a given number of generators
placed in the given size of the system, we fix $L=15$ and $N_{\rm gen} = 49$, and change $L_1$
and $L_2$ systematically.
In Fig.~\ref{fig:L1L2}(a), we fix the size of the inner square to $L_1 = 0$, and vary
$L_2$ from 7 to 15. For $L_2 = 7$, all generators are densely packed in the central part
of the power grid since $S = N_{\rm gen} = 49$, while for $L_2 = L = 15$ generators
are randomly distributed across the whole system. It is observed that the degree of
synchrony is enhanced with $L_2$ for $L_2 = 7, 9,$ and $11$. Interestingly, results for
$L_2 = 13$ and $15$ do not exhibit much difference.
In Fig.~\ref{fig:L1L2}(b), we fix the size of the outer square to $L_2 = L = 15$, and vary
$L_1$ instead. Similarly to Fig.~\ref{fig:L1L2}(a), it is shown that the synchronization
is enhanced as the area available for generators becomes larger.

We next look into the power transmitted through each link from the
expectation that a link can be overloaded if the
link carries too much power and thus it is desirable that the power at each link
is evenly spread over the whole grid.
The power transmitted from node $i$ to node $j$ is given by
\begin{equation}\label{eq:flow}
F_{ij}=-F_{ji}=K\sin(\theta_i-\theta_j),
\end{equation}
with $K$ being the maximum allowed value for transmission.  We measure
$|F_{ij}|$ for each pair of links and construct the normalized histogram
$P(f\equiv |F|/K)$ of the power transmission with bin size $\Delta f =
0.02$.  We measure $P(f)$ at $K=7$ for the cases shown in Figs.~\ref{fig:L1L2}(a) and~\ref{fig:L1L2}(b), and display our results in Figs.~\ref{fig:L1L2}(c) and~\ref{fig:L1L2}(d),
respectively.  The value $K=7$ is chosen since all curves at this value of $K$
exhibit sufficiently large values of $\langle m \rangle$ as shown in
Figs.~\ref{fig:L1L2}(a) and~\ref{fig:L1L2}(b).  In Figs.~\ref{fig:L1L2}(c) and~\ref{fig:L1L2}(d), the color
of each small circle indicates the value of $P(f)$ (see the color bar).
We also measure the average value $\bar f \equiv \sum_f f P(f)$ and the peak
position $f^* \equiv \arg\max P(f)$ and display them in Figs.~\ref{fig:L1L2}(c) and~\ref{fig:L1L2}(d).
It is shown that as the generators are more evenly distributed in the power grid,
i.e., as $L_2$ is increased [Figs.~\ref{fig:L1L2}(a) and~\ref{fig:L1L2}(c)], and as $L_1$ is decreased [Figs.~\ref{fig:L1L2}(b) and~\ref{fig:L1L2}(d)],
the synchronization is enhanced [Figs.~\ref{fig:L1L2}(a) and~\ref{fig:L1L2}(b)] and the transmitted power of links
is more focused on small values [see also the curves for $\bar f$ and $f^*$ in Figs.~\ref{fig:L1L2}(c) and~\ref{fig:L1L2}(d)].

\begin{figure}
\includegraphics[width=0.45\textwidth]{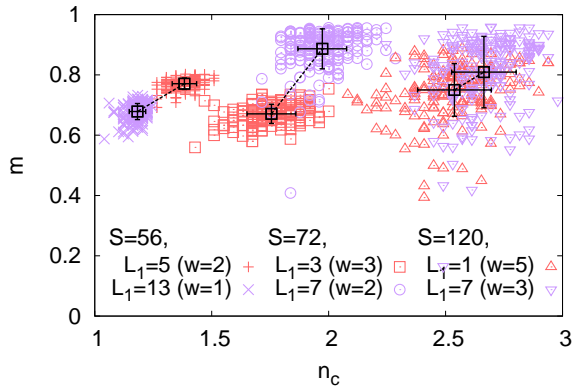}
\caption{(Color online) Scatter plot for $m$ versus $n_c$ at $K=7$
for $L=15$ and $N_{\rm gen} = 49$. To elaborate the effect of $n_c$
for a fixed value of $S = L_2^2 - L_1^2$, we use the parameter values $(L_1, L_2) = (5, 9), (13, 15)$ for $S=56$,
$(3, 9), (7, 11)$ for $S=72$, and $(1, 11), (7, 13)$ for $S=120$.
Positive correlations between $m$ and $n_c$ are seen. The number of samples is 200,
and the ensemble averages of $m$ and $n_c$ are represented as black squares, together
with the error bars.
}
\label{fig:sameS}
\end{figure}

If power plants are packed into a certain narrow region of
the whole system, we believe that the global synchrony may be reduced. As suggested
by the expectation, we measure the number of consumer nodes for which a generator
node provides the power directly.  The average number of direct consumers per
generator is given by
\begin{equation}
\label{eq:pair}
n_c \equiv\frac{1}{N_{\rm gen }}\sum_{i,j}^\prime A_{ij},
\end{equation}
where the primed summation is
for the pair $(i,j)$, with $i$ ($j$) being a generator (a consumer).  In our
two-dimensional square lattice structure for the power grid, the maximum value
of $n_c$ is $4$, the number of nearest neighbors for the square lattice.  In
Figs.~\ref{fig:L1L2}(e) and~\ref{fig:L1L2}(f), we display the scatter plots for $m$
versus $n_c$ at $K=7$, for $L_1 =0$ and $L_2 = 7, 9, 11, 13,$ and $15$ and for $L_2 =15$ and $L_1 = 1, 5, 9, 13$, corresponding to Figs.~\ref{fig:L1L2}(a) and~\ref{fig:L1L2}(c),
and Figs.~\ref{fig:L1L2}(b) and~\ref{fig:L1L2}(d), respectively.  In Fig.~\ref{fig:L1L2}(e), $m$
is increasing with $n_c$, implying that the synchronization is more enhanced
as the number of direct connections between generators and consumers
is increased.

We have shown above that the spatial uniformity for the generator distribution
is closely related to the synchrony of the power grid. In general, as the
region available for generator locations becomes larger, i.e., as the area
$S$ is increased, the synchronization order parameter $m$ becomes larger.
We have also checked the importance of the average number $n_c$ of direct
consumers per generator and have shown that as $n_c$ is increased, so is
$m$. The next question one can ask is, What happens if $n_c$ can be varied
for a given area $S$? To answer the question, we use pairs of
sizes $(L_1, L_2) = (5,9), (13,15)$ for $S=56$,
$(L_1, L_2) = (3,9), (7,11)$ for $S=72$, and
$(L_1, L_2) = (1,11), (7,13)$ for $S=120$, for a system of size $L=15$
with the number of generators $N_{\rm gen}=49$ as for Fig.~\ref{fig:L1L2}
at coupling strength $K = 7$. Figure~\ref{fig:sameS} displays how $m$
changes with $n_c$ for pairs of $(L_1, L_2)$ at fixed values of $S$. Again
we observe that $m$ is an increasing function of $n_c$.

\begin{figure}
\includegraphics[width=0.48\textwidth]{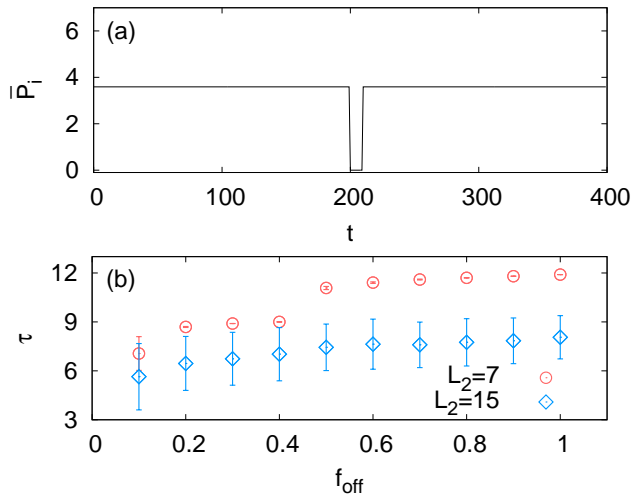}
\caption{(Color online) (a) Temporal failure of a power generator 
is implemented as $\bar{P}_i  =  0$ for $200 \leq t \leq 210$. 
(b) Recovery time $\tau$ versus fraction $f_{\rm off}$
of failed generators (see the text) for $L_{2}=7$ and 15.
We have used $L_{1}=0$ and $K=8$. More uniformly distributed
generators ($L_2 = 15$) exhibit shorter recovery times than
less uniformly distributed generators ($L_2 = 7$).
}
\label{fig:tau}
\end{figure}

We next investigate whether or not the spatial uniformity also yields a good
performance against dynamic perturbation.  It is noteworthy that uniformly
spread plants of small powers can mimic renewable power sources,  which can
occasionally stop operating for various reasons. For example, a wind turbine
may stop producing power if the wind becomes too weak in a certain period of
time.  To mimic such a temporal failure of power generation, we apply the
following scenario: If the $i$-th power generator has a temporal problem, we
change the power production $\bar{P}_i$ as depicted in Fig.~\ref{fig:tau}(a).
That is, $\bar{P}_i$ is set to $0$ only in the time interval $200 \leq t \leq
210$, and before ($t < 200$) and after ($t > 210$) the failure, the power
generation is set to $\bar{P}_i = cP$ as explained in Sec.~\ref{sec:model}. We
apply the failure scenario for a fraction $f_{\rm off}$ of randomly selected
generators.  Note that during the period of failures (i.e., for $200 \leq t
\leq 210$) the power balance condition $\sum_i {\bar P}_i = 0$ is violated, and
thus the power grid as a whole cannot stay in a stationary state. When the
failed plants are back to work at $t=210$, the system is found to recover 
the original level of synchrony after a certain recovery time $\tau$.
We fix $L_1 = 0$ and carry out simulations for $L_2=7$ and 15 
at $K=8$ to investigate the effect of spatial uniformity on the recovery
time $\tau$. Figure~\ref{fig:tau}(b) displays the recovery time $\tau$
(averaged over 200 independent random realizations of plant locations and
failed plants) versus the fraction $f_{\rm off}$ of turned-off power plants for
$L_2 = 7$ and 15.  
It is clearly shown in Fig.~\ref{fig:tau}(b) that 
the more uniformly power sources are distributed, 
the shorter the recovery time $\tau$ is. We thus conclude that spatial
uniformity of the distribution of power plants helps the system
to recover synchrony in shorter times. 

To recap briefly, we have investigated how the uniformity of the
spatial distribution of generators affects the degree of synchrony
and the recovery after a temporal failure of power generation.
As the generators are more uniformly distributed, $n_c$ becomes larger,
leading to an enhancement of the synchronization. It is also observed
that the recovery time becomes shorter for a more uniform distribution of
generators. 
We thus conclude that the spatial uniformity of generator distribution
is beneficial not only for better synchrony but also for better
recovery behavior after a temporal perturbation of power generation.

\begin{figure}
\includegraphics[width=0.48\textwidth]{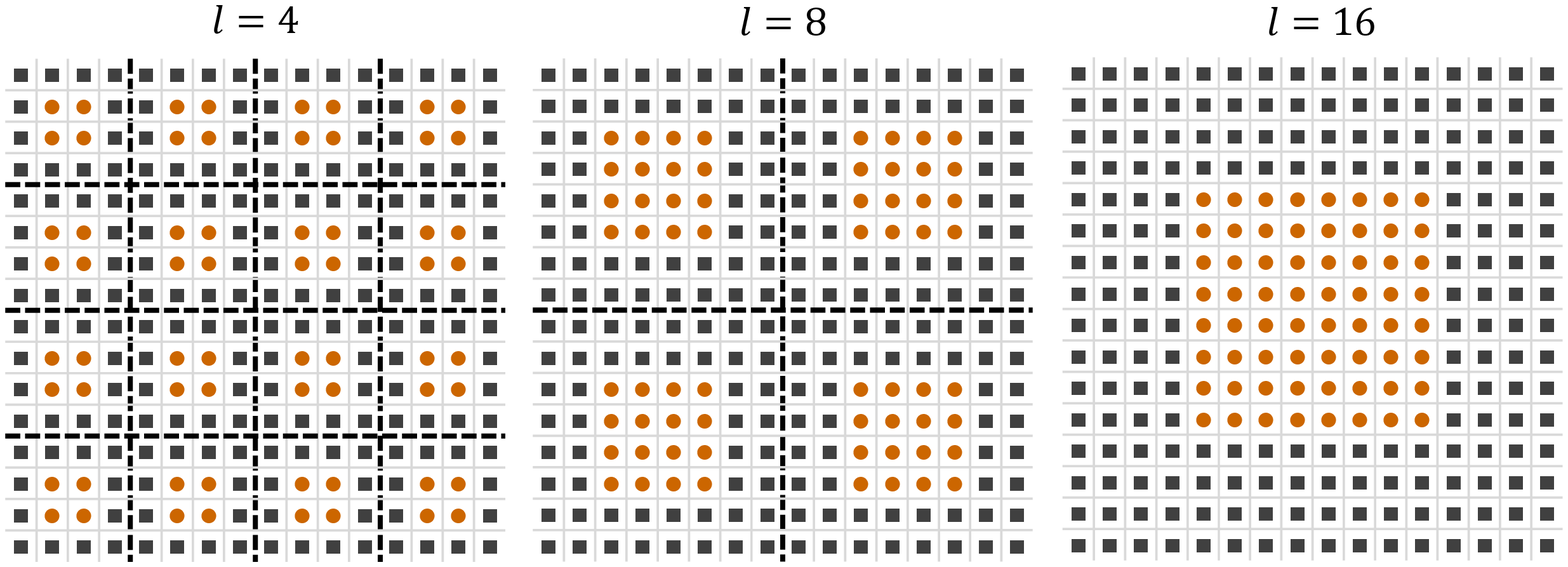}
\caption{(Color online) Method 2: Control of spatial uniformity by imposing
superlattices.  For a two-dimensional power-grid network of the size
$16\times 16$, we placed superlattices of sizes $l \times l$ with $l=4,8,$  and $16$
and generators are packed near the center of each superlattice unit cell.
The total number of generators is $N_{\rm gen} = 64$. Positions of
generators (consumers) are marked by brown circles (black squares) as
in Fig.~\ref{fig:method1}.
}
\label{fig:method2}
\end{figure}

\subsection{Method2: Change of superlattice unit cell}
\label{subsec:super}

We next control the spatial uniformity of the generator distribution with the focus
on the clustering effect.  As depicted in Fig.~\ref{fig:method2}, we place
superlattice structures of different sizes $l$ of unit cells on top of the
original two-dimensional $16 \times 16$ square lattice.  For a given
superlattice of size $l \times l$, we put generators in the central part as
in Fig.~\ref{fig:method2}. The number of generators $N_{\rm gen} = 64$ is fixed
for the system size $N=16 \times 16$ so that the generator density is  1/4,
regardless of the value of $l$.
Note that the smaller the superlattice is, the more uniformly the
spatial distribution of generators becomes.

\begin{figure}
\includegraphics[width=0.47\textwidth]{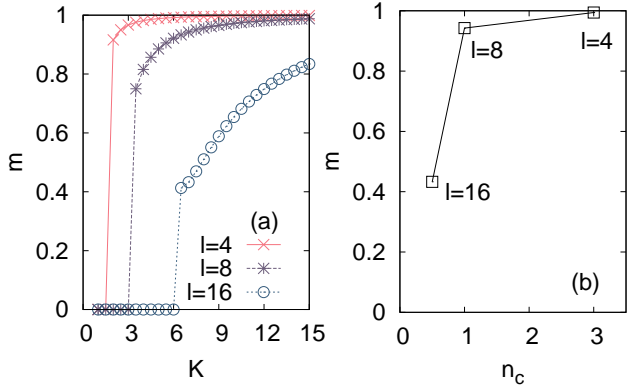}
\caption{(Color online) The spatial uniformity is changed as in Fig.~\ref{fig:method2}.
for $N_{\rm gen} = 64$ and $L=16$.
(a) $m$ versus $K$ for superlattices of size $l=4,8$, and 16.
As $l$ is increased, generators are more clustered, and the level of synchrony becomes
worse.
(b) $m$ versus $n_c$ at $K=7$.
}
\label{fig:cluster}
\end{figure}

In Fig.~\ref{fig:cluster}(a), we plot the order parameter $m$ versus the
coupling strength $K$ for $l=4,8$, and 16. It is clearly shown that as the size
of the superlattice is increased, the global synchrony becomes worse.
Since the positions of generators are given deterministically as shown in
Fig.~\ref{fig:method2}, the sample average is not needed.  We then show in
Fig.~\ref{fig:cluster}(b) how the average number $n_c$ of direct consumers per
generator in Eq.~(\ref{eq:pair}) affects $m$. In accord with our previous observations
for Figs.~\ref{fig:L1L2} and \ref{fig:sameS}, it is clearly shown again that
the spatial uniformity detected by $n_c$ is positively correlated with the
enhancement of the synchrony of the power grid.

\section{Summary and Discussion}
\label{sec:summary}

We have studied how the spatial uniformity in the distribution of power generators
has an influence on the power-grid system.
We have controlled the spatial uniformity, (i) by increasing the number of generators [Fig.~\ref{fig:LNgen}(b)],
and (ii) by increasing the available area for the generators [Figs.~\ref{fig:L1L2}(a) and~\ref{fig:L1L2}(b)],
and have found that the number $n_c$ of power consumers directly
connected to a generator is the key quantity:
The synchronization order parameter $m$  has been found to increase as $n_c$ is increased.
Even when the available area is fixed, $m$ has been shown to be positively correlated with $n_c$
(Fig.~\ref{fig:sameS}). We have also checked the clustering effect of generators for
a given number of generators. It has again been confirmed that $m$ is an increasing function
of $n_c$ also in this case (Fig.~\ref{fig:cluster}). We have also shown that when the global
synchrony is enhanced the probability distribution of the transmitted power of a link tends to have
peaks at small values of the power [Figs.~\ref{fig:L1L2}(c) and~\ref{fig:L1L2}(d)].
This is particularly interesting because the result indicates
that the low values of power transmitted through transmission cables are accompanied by
an enhanced synchrony of the power grid.
It has been also confirmed that for more uniformly distributed generators the
recovery time after a temporal perturbation is shorter, which again
indicates that the spatial uniformity is beneficial (Fig.~\ref{fig:tau}).

In order to better mimic a realistic distribution of power generators, we
consider in Appendix~\ref{App:largeplant} the presence of a few large power
plants along the boundary of the system. 
Also in this case it is found that spatial uniformity of 
small generators tends to ensure better synchrony. 
We still believe that it could be difficult to apply our results 
directly to a real power-grid network.
Nevertheless, we expect that our result will be useful to provide
a guideline for designing a better power-grid structure.

\acknowledgments
We thank Heetae Kim for fruitful discussion and comments. This study was carried out with the support of the Research Program
of Rural Development Administration, Republic of Korea (Project No. PJ01156304).

\appendix
\section{Power transmission through a link}
\label{App:flow}
\begin{figure}[t]
\includegraphics[width=0.48\textwidth]{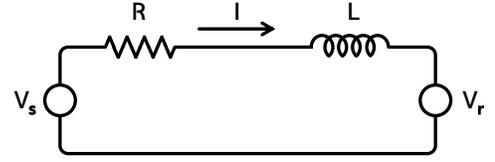}
\caption{A simple transmission line model. The circuit contains a resistor $R$
and an inductor $L$. The limit $R \rightarrow 0$ is taken when we compute the active power
of the circuit.
}
\label{fig:circuit}
\end{figure}

In alternating current circuits at the angular frequency $\Omega$
depicted as a simple transmission line model in Fig.~\ref{fig:circuit},
a sender and a receiver are assigned complex voltages $V_{s}$ and $V_{r}$, respectively.
The complex current is given by $I = (V_s - V_r)/Z$, where the impedance $Z = R + iX_L$
with resistance $R$ and inductive reactance $X_L \equiv \Omega L$.
The complex power $S$ in an ac circuit consists of the active power $P$ and the reactive
power $Q$, i.e., $S=P+iQ=VI^{*}$ with the ac voltage $V$ and the complex conjugate $*$ of
the electric current $I$. Accordingly, the complex powers $S_s$ for the sender and $S_r$
for the receiver are written as
\begin{widetext}
\begin{eqnarray} \label{eq:sender}
S_s &= & P_s + iQ_s =  V_sI^* =\frac{|V_s|^2-V_sV_r^*}{R-iX_L}, \nonumber  \\
S_r &= & P_r + iQ_r =  -V_rI^* =\frac{|V_r|^2-V_rV_s^*}{R-iX_L}.
\end{eqnarray}
We then use $V_{s}= |V_{s}|e^{i\gamma}$ and $V_{r}=|V_{r}|e^{i\delta}$
to get $|V_s|^2-V_sV_r^* =|V_s|^{2}-|V_s||V_r|(\cos\Delta\theta-i\sin\Delta\theta)$ and
       $|V_r|^2-V_rV_s^* =|V_r|^{2}-|V_s||V_r|(\cos\Delta\theta+i\sin\Delta\theta)$,
with $\Delta\theta \equiv \gamma-\delta$, yielding
\begin{eqnarray} \label{eq:finsender}
P_s &=&\frac{R(|V_{s}|^{2}-V_{s}V_{r}\cos\Delta\theta)-X_LV_{s}V_{r}\sin\Delta\theta}{R^2-X_L^2} , \nonumber\\
Q_s &=&-\frac{RV_{s}V_{r}\sin\Delta\theta+X_L(|V_{s}|^{2}-V_{s}V_{r}\cos\Delta\theta)}{R^2-X_L^2}, \nonumber\\
P_r &=&\frac{R(|V_{r}|^{2}-V_{s}V_{r}\cos\Delta\theta)+X_LV_{s}V_{r}\sin\Delta\theta}{R^2-X_L^2}, \nonumber\\
Q_r &=&-\frac{-RV_{s}V_{r}\sin\Delta\theta+X_L(|V_{r}|^{2}-V_{s}V_{r}\cos\Delta\theta)}{R^2-X_L^2}.\\
\end{eqnarray}
The active power $P$ in the limit of zero loss at the resistance ($R = 0$)
is then obtained as $P_s=-P_r=P_{\textrm{max}}\sin\Delta\theta$
with $P_{\textrm{max}}\equiv V_{s}V_{r}/X_{l}$ [see Eq.~(\ref{eq:balance2})].
\end{widetext}

\section{Spatial uniformity with a few large plants along the boundary}
\label{App:largeplant}

\begin{figure}[t]
\includegraphics[width=0.48\textwidth]{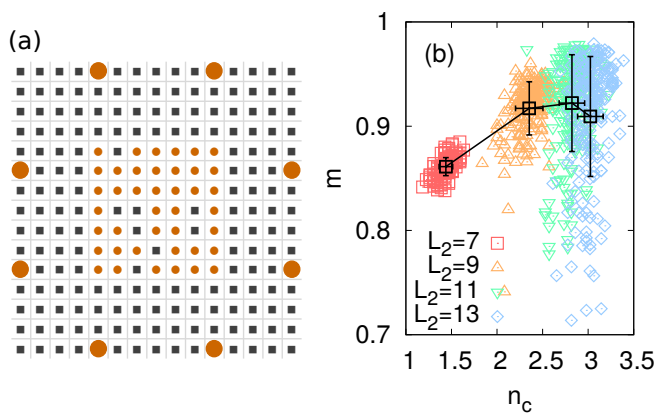}
\caption{(Color online)
(a) For the system discussed in Fig.~\ref{fig:method1}, we place additional
generators of large powers (denoted as larger circles) at fixed positions
along the boundary to mimic a realistic distribution of larger plants along the
coastline. 
(b) Scatter plots of $m$ versus $n_c$ for various vales of $L_2$ 
at fixed values of $L_1 =0$ and $K=7$. Black squares represent averages
over 200 sample runs.}
\label{fig:ncforlarge}
\end{figure}

In reality, we often have larger power plants along the coastline.
To reflect this in our simple model, we introduce 
larger power plants at the boundary as in Fig.~\ref{fig:ncforlarge}(a)
and then control the spatial distribution of only smaller plants in
the inland area (compare with  Sec.~\ref{subsec:spadist}).
We keep $N=N_{\rm gen}+N_{\rm con}=225$, $N_{\rm gen}=49$,
and the demanding power of consumer $\bar{P}_{i}=-P$. 
Among $N_{\rm gen}$ generators, eight generators are regarded as large power
plants with $\bar{P}_i=10P$, and they are located at fixed positions
along the boundary [see Fig.~\ref{fig:ncforlarge}(a)]. 
Other small generators produce the power $\bar{P}_i=c^{\prime}P$
and they are in the middle part of the system.
A constant $c^{\prime}$ is determined from the power
balance condition $\sum_{i}\bar{P}_i=0$. 
The generators with small powers are randomly distributed within the area
formed by various values of $L_{2}$ at fixed $L_{1}=0$.
Scatter-plots of $m$ versus $n_{c}$ at $K=7$ are shown in
Fig.~\ref{fig:ncforlarge}(b). The result demonstrates that the uniform distribution of
small power plants helps to enhance the synchrony with increasing $n_c$, also
in the more realistic case where large power plants are located at fixed
positions along the boundary of the system.

\bibliographystyle{apsrev4-1}

\begin{thebibliography}{19}%
\makeatletter
\providecommand \@ifxundefined [1]{%
 \@ifx{#1\undefined}
}%
\providecommand \@ifnum [1]{%
 \ifnum #1\expandafter \@firstoftwo
 \else \expandafter \@secondoftwo
 \fi
}%
\providecommand \@ifx [1]{%
 \ifx #1\expandafter \@firstoftwo
 \else \expandafter \@secondoftwo
 \fi
}%
\providecommand \natexlab [1]{#1}%
\providecommand \enquote  [1]{``#1''}%
\providecommand \bibnamefont  [1]{#1}%
\providecommand \bibfnamefont [1]{#1}%
\providecommand \citenamefont [1]{#1}%
\providecommand \href@noop [0]{\@secondoftwo}%
\providecommand \href [0]{\begingroup \@sanitize@url \@href}%
\providecommand \@href[1]{\@@startlink{#1}\@@href}%
\providecommand \@@href[1]{\endgroup#1\@@endlink}%
\providecommand \@sanitize@url [0]{\catcode `\\12\catcode `\$12\catcode
  `\&12\catcode `\#12\catcode `\^12\catcode `\_12\catcode `\%12\relax}%
\providecommand \@@startlink[1]{}%
\providecommand \@@endlink[0]{}%
\providecommand \url  [0]{\begingroup\@sanitize@url \@url }%
\providecommand \@url [1]{\endgroup\@href {#1}{\urlprefix }}%
\providecommand \urlprefix  [0]{URL }%
\providecommand \Eprint [0]{\href }%
\providecommand \doibase [0]{http://dx.doi.org/}%
\providecommand \selectlanguage [0]{\@gobble}%
\providecommand \bibinfo  [0]{\@secondoftwo}%
\providecommand \bibfield  [0]{\@secondoftwo}%
\providecommand \translation [1]{[#1]}%
\providecommand \BibitemOpen [0]{}%
\providecommand \bibitemStop [0]{}%
\providecommand \bibitemNoStop [0]{.\EOS\space}%
\providecommand \EOS [0]{\spacefactor3000\relax}%
\providecommand \BibitemShut  [1]{\csname bibitem#1\endcsname}%
\let\auto@bib@innerbib\@empty
\bibitem [{\citenamefont {Kolmos}\ and\ \citenamefont
  {Davis}(2007)}]{circadian}%
  \BibitemOpen
  \bibfield  {author} {\bibinfo {author} {\bibfnamefont {E.}~\bibnamefont
  {Kolmos}}\ and\ \bibinfo {author} {\bibfnamefont {S.~J.}\ \bibnamefont
  {Davis}},\ }\href@noop {} {\bibfield  {journal} {\bibinfo  {journal} {Curr.
  Biol.}\ }\textbf {\bibinfo {volume} {17}},\ \bibinfo {pages} {R808} (\bibinfo
  {year} {2007})}\BibitemShut {NoStop}%
\bibitem [{\citenamefont {Fisher}\ \emph {et~al.}(2005)\citenamefont {Fisher},
  \citenamefont {v.~E.~Boas}, \citenamefont {Blume}, \citenamefont {Elger},
  \citenamefont {Genton}, \citenamefont {P.Lee},\ and\ \citenamefont
  {Engel}}]{epil}%
  \BibitemOpen
  \bibfield  {author} {\bibinfo {author} {\bibfnamefont {R.~S.}\ \bibnamefont
  {Fisher}}, \bibinfo {author} {\bibfnamefont {W.}~\bibnamefont {v.~E.~Boas}},
  \bibinfo {author} {\bibfnamefont {W.}~\bibnamefont {Blume}}, \bibinfo
  {author} {\bibfnamefont {C.}~\bibnamefont {Elger}}, \bibinfo {author}
  {\bibfnamefont {P.}~\bibnamefont {Genton}}, \bibinfo {author} {\bibnamefont
  {P.Lee}}, \ and\ \bibinfo {author} {\bibfnamefont {J.}~\bibnamefont
  {Engel}},\ }\href@noop {} {\bibfield  {journal} {\bibinfo  {journal}
  {Epilepsia}\ }\textbf {\bibinfo {volume} {46}},\ \bibinfo {pages} {470}
  (\bibinfo {year} {2005})}\BibitemShut {NoStop}%
\bibitem [{\citenamefont {Strogatz}\ \emph {et~al.}(2005)\citenamefont
  {Strogatz}, \citenamefont {Abrams}, \citenamefont {McRobie}, \citenamefont
  {Eckhardt},\ and\ \citenamefont {Ott}}]{bridge}%
  \BibitemOpen
  \bibfield  {author} {\bibinfo {author} {\bibfnamefont {S.~H.}\ \bibnamefont
  {Strogatz}}, \bibinfo {author} {\bibfnamefont {D.~M.}\ \bibnamefont
  {Abrams}}, \bibinfo {author} {\bibfnamefont {A.}~\bibnamefont {McRobie}},
  \bibinfo {author} {\bibfnamefont {B.}~\bibnamefont {Eckhardt}}, \ and\
  \bibinfo {author} {\bibfnamefont {E.}~\bibnamefont {Ott}},\ }\href@noop {}
  {\bibfield  {journal} {\bibinfo  {journal} {Nature (London)}\ }\textbf
  {\bibinfo {volume} {438}},\ \bibinfo {pages} {43} (\bibinfo {year}
  {2005})}\BibitemShut {NoStop}%
\bibitem [{\citenamefont {Silverberg}\ \emph {et~al.}(2013)\citenamefont
  {Silverberg}, \citenamefont {Bierbaum}, \citenamefont {Sethna},\ and\
  \citenamefont {Cohen}}]{social1}%
  \BibitemOpen
  \bibfield  {author} {\bibinfo {author} {\bibfnamefont {J.~L.}\ \bibnamefont
  {Silverberg}}, \bibinfo {author} {\bibfnamefont {M.}~\bibnamefont
  {Bierbaum}}, \bibinfo {author} {\bibfnamefont {J.~P.}\ \bibnamefont
  {Sethna}}, \ and\ \bibinfo {author} {\bibfnamefont {I.}~\bibnamefont
  {Cohen}},\ }\href@noop {} {\bibfield  {journal} {\bibinfo  {journal} {Phys.
  Rev. Lett.}\ }\textbf {\bibinfo {volume} {110}},\ \bibinfo {pages} {228701}
  (\bibinfo {year} {2013})}\BibitemShut {NoStop}%
\bibitem [{\citenamefont {Hoang}\ \emph {et~al.}(2015)\citenamefont {Hoang},
  \citenamefont {Jo},\ and\ \citenamefont {Hong}}]{social2}%
  \BibitemOpen
  \bibfield  {author} {\bibinfo {author} {\bibfnamefont {D.-T.}\ \bibnamefont
  {Hoang}}, \bibinfo {author} {\bibfnamefont {J.}~\bibnamefont {Jo}}, \ and\
  \bibinfo {author} {\bibfnamefont {H.}~\bibnamefont {Hong}},\ }\href@noop {}
  {\bibfield  {journal} {\bibinfo  {journal} {Phys. Rev. E}\ }\textbf {\bibinfo
  {volume} {91}},\ \bibinfo {pages} {032135} (\bibinfo {year}
  {2015})}\BibitemShut {NoStop}%
\bibitem [{\citenamefont {Filatrella}\ \emph {et~al.}(2008)\citenamefont
  {Filatrella}, \citenamefont {Nielsen},\ and\ \citenamefont
  {Pedersen}}]{epjb}%
  \BibitemOpen
  \bibfield  {author} {\bibinfo {author} {\bibfnamefont {G.}~\bibnamefont
  {Filatrella}}, \bibinfo {author} {\bibfnamefont {A.~H.}\ \bibnamefont
  {Nielsen}}, \ and\ \bibinfo {author} {\bibfnamefont {N.~F.}\ \bibnamefont
  {Pedersen}},\ }\href@noop {} {\bibfield  {journal} {\bibinfo  {journal} {Eur.
  Phys. J. B}\ }\textbf {\bibinfo {volume} {61}},\ \bibinfo {pages} {485}
  (\bibinfo {year} {2008})}\BibitemShut {NoStop}%
\bibitem [{\citenamefont {Louzada}\ \emph {et~al.}(2012)\citenamefont
  {Louzada}, \citenamefont {Ara\'ujo}, \citenamefont {Jr.},\ and\ \citenamefont
  {Herrmann}}]{suppress}%
  \BibitemOpen
  \bibfield  {author} {\bibinfo {author} {\bibfnamefont {V.~H.~P.}\
  \bibnamefont {Louzada}}, \bibinfo {author} {\bibfnamefont {N.~A.~M.}\
  \bibnamefont {Ara\'ujo}}, \bibinfo {author} {\bibfnamefont {J.~S.~A.}\
  \bibnamefont {Jr.}}, \ and\ \bibinfo {author} {\bibfnamefont {H.~J.}\
  \bibnamefont {Herrmann}},\ }\href@noop {} {\bibfield  {journal} {\bibinfo
  {journal} {Sci. Rep.}\ }\textbf {\bibinfo {volume} {2}},\ \bibinfo {pages}
  {658} (\bibinfo {year} {2012})}\BibitemShut {NoStop}%
\bibitem [{\citenamefont {Mazloomzadeh}\ \emph {et~al.}(2012)\citenamefont
  {Mazloomzadeh}, \citenamefont {Salehi},\ and\ \citenamefont
  {Mohammed}}]{gridsynch1}%
  \BibitemOpen
  \bibfield  {author} {\bibinfo {author} {\bibfnamefont {A.}~\bibnamefont
  {Mazloomzadeh}}, \bibinfo {author} {\bibfnamefont {V.}~\bibnamefont
  {Salehi}}, \ and\ \bibinfo {author} {\bibfnamefont {O.}~\bibnamefont
  {Mohammed}},\ }in\ \href@noop {} {\emph {\bibinfo {booktitle} {2012 IEEE PES
  Innovative Smart Grid Technologies (ISGT)}}}\ (\bibinfo  {publisher} {IEEE},\
  \bibinfo {address} {Washington, DC},\
  \bibinfo {year} {2012})\
  p.~\bibinfo {pages} {1}\BibitemShut {NoStop}%
\bibitem [{\citenamefont {Croft}\ and\ \citenamefont
  {Summers}(1987)}]{gridsynch2}%
  \BibitemOpen
  \bibfield  {author} {\bibinfo {author} {\bibfnamefont {T.}~\bibnamefont
  {Croft}}\ and\ \bibinfo {author} {\bibfnamefont {W.}~\bibnamefont
  {Summers}},\ }\href@noop {} {\emph {\bibinfo {title} {American Electricans'
  Handbook, Eleventh Edition}}}\ (\bibinfo  {publisher} {McGraw-Hill},\
  \bibinfo {address} {New York},\ \bibinfo {year} {1987})\BibitemShut {NoStop}%
\bibitem [{\citenamefont {Nishikawa}\ and\ \citenamefont
  {Motter}(2015)}]{motternjp}%
  \BibitemOpen
  \bibfield  {author} {\bibinfo {author} {\bibfnamefont {T.}~\bibnamefont
  {Nishikawa}}\ and\ \bibinfo {author} {\bibfnamefont {A.~E.}\ \bibnamefont
  {Motter}},\ }\href@noop {} {\bibfield  {journal} {\bibinfo  {journal} {New J.
  Phys.}\ }\textbf {\bibinfo {volume} {17}},\ \bibinfo {pages} {015012}
  (\bibinfo {year} {2015})}\BibitemShut {NoStop}%
\bibitem [{\citenamefont {Menck}\ \emph {et~al.}(2013)\citenamefont {Menck},
  \citenamefont {Heitzig}, \citenamefont {Marwan},\ and\ \citenamefont
  {Kruths}}]{kruths}%
  \BibitemOpen
  \bibfield  {author} {\bibinfo {author} {\bibfnamefont {P.~J.}\ \bibnamefont
  {Menck}}, \bibinfo {author} {\bibfnamefont {J.}~\bibnamefont {Heitzig}},
  \bibinfo {author} {\bibfnamefont {N.}~\bibnamefont {Marwan}}, \ and\ \bibinfo
  {author} {\bibfnamefont {J.}~\bibnamefont {Kruths}},\ }\href@noop {}
  {\bibfield  {journal} {\bibinfo  {journal} {Nat. Phys.}\ }\textbf {\bibinfo
  {volume} {9}},\ \bibinfo {pages} {89} (\bibinfo {year} {2013})}\BibitemShut
  {NoStop}%
\bibitem [{\citenamefont {Kim}\ \emph {et~al.}(2016)\citenamefont {Kim},
  \citenamefont {Lee},\ and\ \citenamefont {Holme}}]{hkim1}%
  \BibitemOpen
  \bibfield  {author} {\bibinfo {author} {\bibfnamefont {H.}~\bibnamefont
  {Kim}}, \bibinfo {author} {\bibfnamefont {S.~H.}\ \bibnamefont {Lee}}, \ and\
  \bibinfo {author} {\bibfnamefont {P.}~\bibnamefont {Holme}},\ }\href@noop {}
  {\bibfield  {journal} {\bibinfo  {journal} {Phys. Rev. E}\ }\textbf {\bibinfo
  {volume} {93}},\ \bibinfo {pages} {062318} (\bibinfo {year}
  {2016})}\BibitemShut {NoStop}%
\bibitem [{\citenamefont {Li}\ and\ \citenamefont {Wong}()}]{msf}%
  \BibitemOpen
  \bibfield  {author} {\bibinfo {author} {\bibfnamefont {B.}~\bibnamefont
  {Li}}\ and\ \bibinfo {author} {\bibfnamefont {K.~Y.~M.}\ \bibnamefont
  {Wong}},\ }\href@noop {} {}\bibinfo {note} {{arXiv:1607.04509}}\BibitemShut
  {NoStop}%
\bibitem [{\citenamefont {Rohden}\ \emph {et~al.}(2014)\citenamefont {Rohden},
  \citenamefont {Sorge}, \citenamefont {Witthaut},\ and\ \citenamefont
  {Timme}}]{impact}%
  \BibitemOpen
  \bibfield  {author} {\bibinfo {author} {\bibfnamefont {M.}~\bibnamefont
  {Rohden}}, \bibinfo {author} {\bibfnamefont {A.}~\bibnamefont {Sorge}},
  \bibinfo {author} {\bibfnamefont {D.}~\bibnamefont {Witthaut}}, \ and\
  \bibinfo {author} {\bibfnamefont {M.}~\bibnamefont {Timme}},\ }\href@noop {}
  {\bibfield  {journal} {\bibinfo  {journal} {Chaos}\ }\textbf {\bibinfo
  {volume} {24}},\ \bibinfo {pages} {013123} (\bibinfo {year}
  {2014})}\BibitemShut {NoStop}%
\bibitem [{\citenamefont {Pinto}\ and\ \citenamefont {Saa}(2016)}]{optimize}%
  \BibitemOpen
  \bibfield  {author} {\bibinfo {author} {\bibfnamefont {R.~S.}\ \bibnamefont
  {Pinto}}\ and\ \bibinfo {author} {\bibfnamefont {A.}~\bibnamefont {Saa}},\
  }\href@noop {} {\bibfield  {journal} {\bibinfo  {journal} {Physica A}\
  }\textbf {\bibinfo {volume} {463}},\ \bibinfo {pages} {77} (\bibinfo {year}
  {2016})}\BibitemShut {NoStop}%
\bibitem [{\citenamefont {Rohden}\ \emph {et~al.}(2012)\citenamefont {Rohden},
  \citenamefont {Sorge}, \citenamefont {Timme},\ and\ \citenamefont
  {Witthaut}}]{decentral}%
  \BibitemOpen
  \bibfield  {author} {\bibinfo {author} {\bibfnamefont {M.}~\bibnamefont
  {Rohden}}, \bibinfo {author} {\bibfnamefont {A.}~\bibnamefont {Sorge}},
  \bibinfo {author} {\bibfnamefont {M.}~\bibnamefont {Timme}}, \ and\ \bibinfo
  {author} {\bibfnamefont {D.}~\bibnamefont {Witthaut}},\ }\href@noop {}
  {\bibfield  {journal} {\bibinfo  {journal} {Phys. Rev. Lett.}\ }\textbf
  {\bibinfo {volume} {109}},\ \bibinfo {pages} {064101} (\bibinfo {year}
  {2012})}\BibitemShut {NoStop}%
\bibitem [{\citenamefont {Witthaut}\ and\ \citenamefont
  {Timme}(2012)}]{braess}%
  \BibitemOpen
  \bibfield  {author} {\bibinfo {author} {\bibfnamefont {D.}~\bibnamefont
  {Witthaut}}\ and\ \bibinfo {author} {\bibfnamefont {M.}~\bibnamefont
  {Timme}},\ }\href@noop {} {\bibfield  {journal} {\bibinfo  {journal} {New J.
  Phys.}\ }\textbf {\bibinfo {volume} {14}},\ \bibinfo {pages} {083036}
  (\bibinfo {year} {2012})}\BibitemShut {NoStop}%
\bibitem [{\citenamefont {Press}\ \emph {et~al.}(1992)\citenamefont {Press},
  \citenamefont {Teukolsky}, \citenamefont {Vetterling},\ and\ \citenamefont
  {Flannery}}]{Press}%
  \BibitemOpen
  \bibfield  {author} {\bibinfo {author} {\bibfnamefont {W.~H.}\ \bibnamefont
  {Press}}, \bibinfo {author} {\bibfnamefont {S.~A.}\ \bibnamefont
  {Teukolsky}}, \bibinfo {author} {\bibfnamefont {W.~T.}\ \bibnamefont
  {Vetterling}}, \ and\ \bibinfo {author} {\bibfnamefont {B.~P.}\ \bibnamefont
  {Flannery}},\ }\href@noop {} {\emph {\bibinfo {title} {Numerical Recipes in
  C: The Art of Scientific Computing}}},\ \bibinfo {edition} {2nd}\ ed.\
  (\bibinfo  {publisher} {Cambridge University Press},\ \bibinfo {address} {New
  York, USA},\ \bibinfo {year} {1992})\BibitemShut {NoStop}%
\bibitem [{\citenamefont {Hong}\ \emph {et~al.}(2004)\citenamefont {Hong},
  \citenamefont {Park},\ and\ \citenamefont {Choi}}]{localcouple}%
  \BibitemOpen
  \bibfield  {author} {\bibinfo {author} {\bibfnamefont {H.}~\bibnamefont
  {Hong}}, \bibinfo {author} {\bibfnamefont {H.}~\bibnamefont {Park}}, \ and\
  \bibinfo {author} {\bibfnamefont {M.~Y.}\ \bibnamefont {Choi}},\ }\href@noop
  {} {\bibfield  {journal} {\bibinfo  {journal} {Phys. Rev. E}\ }\textbf
  {\bibinfo {volume} {70}},\ \bibinfo {pages} {045204} (\bibinfo {year}
  {2004})}\BibitemShut {NoStop}%
\end{thebibliography}

%

\end{document}